\documentstyle[seceq,letter]{ptptex}

\title{
Relativistic corrections to astrometric shifts due to 
gravitational microlensing
}

\author{
Geraint F.  {\sc Lewis}$^{1,}$\footnote{E-mail address:
gfl@aaoepp.aao.gov.au} 
and
X. Rosalind {\sc Wang}$^{2,}$\footnote{E-mail address: 
xywang@mail.usyd.edu.au}
}

\inst{
$^1$Anglo-Australian Observatory, PO Box 296, Epping, NSW 1710, Australia
\\
$^2$Dept. of Mechanical and Mechatronics, Uni. of Sydney, 
NSW 2006, Australia
}

\recdate{
}

\abst{
Gravitational microlensing has proved  to be a versatile astrophysical
tool.   Recently, the  question of  whether higher  order relativistic
corrections  can influence the  observable properties  of microlensing
has been addressed. This letter extends earlier studies, investigating
higher   order  relativistic  terms   in  astrometric   shifts  during
microlensing  events,  revealing that  they  produce  no effects.  The
magnitude  of   the  relativistic   corrections  is  examined   in  an
astrophysical context and is found to be negligible.}

\begin{document}

\maketitle

\section{Introduction}\label{introduction}
Over the  past decade, the field of  gravitational microlensing within
the Local Group has grown rapidly. Several hundred microlensing events
have   been   detected   by   large  monitoring   surveys,   including
MACHO~\cite{rf:1}, EROS~\cite{rf:2}  and OGLE~\cite{rf:3}.  These have
provided important  clues to the  mass distribution of the  Galaxy, as
well  as   unique  view  of  detailed  surface   features  of  distant
stars~\cite{rf:4,rf:5,rf:6}.  Microlensing also holds great promise in
the  identification of  planetary systems  beyond the  range  of other
observational   approaches~\cite{rf:7,rf:8,rf:9}.    The  details   of
gravitational    microlensing   are    given    in   several    recent
reviews~\cite{rf:10} and will not be reproduced here.

The  standard  formalism  employed   in  the  study  of  gravitational
microlensing considered the relativistic deflection up to first order,
assuming  higher order  correction  are small  and  can be  neglected.
Recently,  Ebina, Osuga, Asada  \& Kasai~\cite{rf:11}  have questioned
this assumption  and considered second  order relativistic corrections
to the  currently employed  formalism.  Intriguingly, they  found that
while both the image positions  and magnifications are modified by the
addition of higher  order terms, the sum of  the magnifications, which
is  the  observed quantity  during  the  photometric  monitoring of  a
microlensing event, does not suffer any second order correction.

Here,  the   study  of  Ebina  et  al.~\cite{rf:11}   is  extended  to
investigate the magnitude of the second order relativistic corrections
to    the   image    astrometric   shift    during    a   microlensing
event~\cite{rf:12}.

\section{Background}\label{background}
All angles are normalized to the natural scale length of gravitational
microlensing, the angular Einstein radius;
\begin{equation}
        \theta_{E} = \sqrt{ \frac{4GM}{c^2} \frac{D_{ds}} { D_{d} D_{s} } }
\end{equation}
where $M$ is the mass  of the compact microlensing object and $D_{s}$,
$D_{d}$ and  $D_{ds}$ represents the relative  observer, deflector and
source distances~\cite{rf:13}.  For source  at a angle $\theta_s$ from
a lensing mass, two images are produced at angles;
\begin{equation} 
     \theta_{1,2} = \frac{1}{2} (\theta_{s} \pm \sqrt{\theta_{s} ^ 2 + 4 })
     \label{eq:psn1}
\end{equation}
For  stellar mass  lenses within  the Local  Group, the  separation of
these  images is  typically milliarcseconds,  below the  resolution of
current  optical observations.  These  images are,  however, magnified
with respect to the unlensed source, the magnifications given by;
\begin{equation}
        \mu_{1,2} = 
                   \frac { \theta_{s}^2 + 2 } 
                         {2 \theta_{s} \sqrt{ \theta_{s}^2 + 4 } }
                   \pm \frac{1}{2}\ ,
\ \ \ 
         \mu_{tot} = \mu_{1} + \mu_{2} 
             = \frac {\theta_{s}^2 + 2} 
                     {\theta_{s} \sqrt{\theta_{s}^2 + 4} }
         \label{eq:tot-amp}
\end{equation}
The magnification can be  substantial, resulting in the characteristic
`bell-shaped' light curve observed during a gravitational microlensing
event.

Combining  the  individual  image  positions and  magnifications,  the
centroid of  the light  distribution of a  gravitationally microlensed
source is given by;
\begin{equation}
        \theta_{cent} = 
           \frac { \theta_{1} \mu_{1} +  \theta_{2} \mu_{2} }
                 {\mu_{1} + \mu_{2}}
        \label{eq:centroid}
\end{equation}
During  a microlensing  event, therefore,  as a  source  brightens and
fades     it     should     also     exhibit     a     shift     light
centroid~\cite{rf:14,rf:15}. This  will be  small, again the  order of
milliarcseconds, but such centroid shifts are detectable with the next
generation  of  space  interferometers~\cite{rf:16,rf:17}, which  will
have     an     accuracy      of     microarcseconds     (see     {\tt
http://sim.jpl.nasa.gov/}).

\section{Second Order Lensing Correction}\label{2ndOrder}
In the weak gravitational  fields of typical astrophysical regimes, it
is usual to consider the influence of general relativity to low order.
The question of higher  order corrections to the gravitational lensing
have  been considered previously~\cite{rf:18},  and recently  Ebina et
al.~\cite{rf:11} examined  the magnitude of  second order relativistic
effects to the gravitational microlensing equations.  Considering this
higher order term to be a perturbation to the solutions of the lensing
equation, demonstrated the image positions become
\begin{equation}
\theta'_{1,2}   =    \theta_{1,2}   +   \frac{\lambda}{   \theta_{1,2}
\sqrt{\theta_{s} ^2  + 4} } + {\cal  O} (\lambda^2) \label{eq:psn-2nd}\ ,
\ \ \
           \lambda = \frac{15 \pi D_{s} \theta_{E}}{64D_{ds}} 
           \label{eq:lambda}
\end{equation}
Here, $\lambda$  is a dimensionless parameter, the  magnitude of which
is considered  in Section~5.  Similarly,  the inclusion of  the second
order term modifies the image magnifications, such that
\begin{equation}
        \mu'_{1,2} = \mu_{1,2} 
             \mp \lambda (\theta_{s}^2 + 4)^{-3/2} 
             + {\cal O} (\lambda^2) \ , 
\ \ \ \
        \mu'_{tot} = \frac { \theta_{s}^2 + 2 } 
                           { \theta_{s} \sqrt{ \theta_{s}^2 + 4 } }
                     + {\cal O}(\lambda^2)
	= \mu_{tot}
\label{eq:mag}
\end{equation}
Surprisingly,  while the  magnification  of the  individual images  is
modified  by  the  inclusion  of  the second  order  term,  the  total
magnification is  not, demonstrating that the inclusion  of the higher
order relativistic correction produces no modification of microlensing
light curves.

\section{Astrometric Shifts}\label{shift}
While  the total  magnification  is independent  of  the second  order
corrections,  the   previous  section  demonstrates   that  the  image
positions      and       magnifications      change.       Considering
Equation~\ref{eq:centroid}, these  changes can potentially  modify the
path of the  image centroid during a microlensing  event and result in
detectable consequences of the higher order relativistic corrections.

Again defining the image centroid to be
(Equation~\ref{eq:centroid})
\begin{equation}
        \theta'_{cent} = 
          \frac
            {\theta'_{1} \mu'_{1}  +  \theta'_{2} \mu'_{2}  }
            {\mu'_{1} + \mu'_{2}}
        \label{eq:cent-2nd}
\end{equation}
and  expanding  this  expression using  Equations~\ref{eq:lambda}  and
\ref{eq:mag}, this becomes
\begin{equation}
        \theta'_{cent} = \theta_{cent} + 
        \left(  
          \frac{ \frac{\lambda}{(\theta_{s}^2 + 4)^{3/2}} 
                       \left(\theta_{2}-\theta_{1}\right)
                 + \frac{\lambda}{(\theta_{s}^2 + 4)^{1/2}} 
                       \left(\frac{\mu_{1}}{\theta_{1}} 
                         + \frac{\mu_{2}}{\theta_{2}}
                       \right)
                 + {\cal O}(\lambda^2)
                        }
               {\mu_{1} + \mu_{2}}
        \right)
        \label{eq:cent-2nd-exp}    
\end{equation}
A  closer examination  reveals that  the  first two  terms within  the
brackets  are the  equivalent, but  of opposite  sign, and  so cancel.
Hence, when accounting for  second order relativistic corrections, the
centroid   shift   is   $\theta'_{cent}   =  \theta_{cent}   +   {\cal
O}(\lambda^2)$, and, like the total magnification, the position of the
image  centroid is  independent of  these higher  order  terms. Hence,
consideration  of  the second  order  relativistic  correction to  the
equations of  gravitational microlensing produces no  deviation to the
observational properties available to us during a microlensing event.

\section{How big is $\lambda$ ?}\label{lambda}
The previous  sections have demonstrated  how the inclusion  of second
order  relativistic  corrections  to  the formalism  of  gravitational
microlensing produces  no observable  consequences to the  total image
magnification and  centroid shift. With  Equations~\ref{eq:lambda} and
\ref{eq:mag}, the  individual image magnifications  and positions are,
however, modified by the higher order terms. It is prudent, therefore,
to  ask that,  if  observational techniques  improve  to allowing  the
resolution   of   the  individual   microlensed   images,  would   the
consequences of the relativistic corrections be then observable?  This
question depends entirely on the magnitude of the parameter $\lambda$.

In  considering Equation~\ref{eq:lambda}, the  numerical value  of the
second order  parameter $\lambda$ for a  typical microlensing scenario
within the Local Group is;
\begin{equation} 
 \lambda  =  3.6   \times  10^{-9}  \frac{1}{\left[  q  (   1  -  q  )
	\right]^\frac{1}{2}}
	\left[\frac{M}{M_\odot}\right]^\frac{1}{2}
	\left[\frac{D_s}{8kpc}\right]^{-\frac{1}{2}}
\end{equation}
where $q={D_d}/{D_s}$, the ratio of  the distances to the lens and the
source, $M_\odot$ is the mass of the Sun and $8kpc$ is distance to the
Galactic centre.   It must be  remembered that the units  presented in
this paper are normalized with respect to the angular Einstein radius,
hence the deviations due  to the second order relativistic corrections
are of order  $\lambda \times \theta_E$.  Given the  above, it must be
concluded that  the deviations of  the individual image  positions and
magnification due  to such a correction  is astrophysically negligible
and beyond detection.

\section{Conclusion}\label{conclusion}
This  paper has examined  the influence  of second  order relativistic
corrections   on  the   magnitude   of  astrometic   shift  during   a
gravitational  microlensing  event.   It  is  demonstrated  that  this
quantity, like the total  magnification, is independent of this higher
order  term. It is  also shown  that the  second order  correction for
gravitational  microlensing  within   the  Local  Group  is  typically
extremely small, and hence has negligible astrophysical impact.

\section*{Acknowledgements}
X. R.  Wang  would like to thank the  Anglo-Australian Observatory for
the hospitality and support of the research conducted as a part of the
AAO's Summer Scholarship Program.

\end{document}